\documentclass[12pt]{article}

\usepackage{amssymb}
\usepackage{amsmath}
\usepackage{amscd}
\usepackage{latexsym}
\usepackage{graphicx}

\usepackage{enumerate}

\usepackage{cite}

\newcommand{\be}{\begin{equation}}
\newcommand{\ee}{\end{equation}}

\newcommand{\dlt}{\delta}

\newcommand{\br}{{\bf r}}
\newcommand{\bk}{{\bf k}}

\newcommand{\ba}{{\bf a}}

\newcommand{\vp}{\varphi}
\newcommand{\ep}{\varepsilon}
\newcommand{\al}{\alpha}
\newcommand{\ra}{\rightarrow}
\newcommand{\sgm}{\sigma}

\newcommand{\om}{\omega}

\newcommand{\dgr}{\dagger}

\newcommand{\cH}{{\cal H}}

\newcommand{\cL}{{\cal L}}
\newcommand{\rgl}{\rangle}
\newcommand{\lgl}{\langle}

\begin{document}

\begin{center}
{\Large {\bf Correlation measure for statistical systems} \\ [5mm]

V.I. Yukalov$^{1,2}$ and E.P. Yukalova$^3$ } \\ [3mm]

{\it $^1$Bogolubov Laboratory of Theoretical Physics, \\
Joint Institute for Nuclear Research, Dubna 141980, Russia \\ [2mm]

$^2$Instituto de Fisica de S\~ao Carlos, Universidade de S\~ao Paulo, \\
CP 369, S\~ao Carlos 13560-970, S\~ao Paulo, Brazil \\ [2mm]

$^3$Laboratory of Information Technologies, \\
Joint Institute for Nuclear Research, Dubna 141980, Russia } \\ [3mm]

{\bf E-mails}: yukalov@theor.jinr.ru, yukalova@theor.jinr.ru

\end{center}

\vskip 2cm

\begin{abstract}

A measure of correlations for statistical systems is advanced, called correlation index.
The measure is general, being applicable to quantum and classical, equilibrium and 
nonequilibrium statistical systems. After formulating the general notion of correlation 
indices for arbitrary trace-class operators, the calculation of the correlation indices are     
illustrated by several examples of density operators and reduced density matrices. The 
computation of the correlation indices for equilibrium systems is exemplified for bosonic 
systems with the Bose-Einstein condensate, superconducting systems, and for spin systems. 
Correlation indices for nonequilibrium systems are considered for nonlinear dynamics of 
spin and pseudospin systems.     
    
\end{abstract}

\vskip 1cm
{\parindent=0pt
{\bf Keywords}: Statistical systems, Correlation indices, Bose-Einstein condensation, Spin 
systems, Nonlinear spin dynamics}

\newpage

\section{Introduction}

The majority of systems around us are composite, being composed of several parts. Thus even
the simplest quantum systems can contain a number of parts or bits. Statistical systems by
default are composite, being formed by many atoms or molecules. The system properties depend
on the strength of correlations between the system parts. There exists an old standing 
problem of how to measure the level of correlations between the parts of the system. This 
problem should not be confused with the study of correlations in series of numerical statistical 
data, where one usually considers the familiar Pearson coefficient 
\cite{Aitken_1,Croxton_2,Dietrich_3}. The main objects of our investigation are statistical 
systems that are, by definition, composite, being composed of many parts, such as atoms or 
molecules. 
  
Qualitatively, the type of correlations can be judged by the qualitative behavior of 
correlation functions. Thus, an exponentially decaying correlation function signifies 
short-range correlations, while an algebraically diminishing function displays long-range 
correlations. This, however, is a qualitative classification. But how would it be possible 
to make a quantitative comparison of correlations, reducing the problem to the comparison of 
just numbers?

In the present paper, we suggest an approach for a quantitative measure of correlations. This
approach is general, being applicable to quantum and classical systems, to equilibrium as well 
as nonequilibrium systems. In Sec. 2, the general definition is given for the correlation 
indices of arbitrary trace-class operators acting on tensor-product spaces. The introduced 
measure is positive-defined. Examples of correlation indices for quantum states are given. 
In Sec. 3, correlation indices are constructed on the basis of reduced density matrices. 
Section 4 presents a detailed characterization of correlation indices for trapped gases with 
the Bose-Einstein condensate and Sec. 5, for superconducting systems. In Sec. 6, we compute 
correlation indices for spin systems. And in Sec. 7, we demonstrate the temporal behavior 
of correlation indices for spin and pseudospin nonlinear dynamics of some nonequilibrium 
systems. Section 8 concludes.

\section{Correlation indices for operators}

The notion of correlation indices can be introduced for operators acting on a composite 
Hilbert space in the form of a tensor product
\be
\label{1}
 \cH \; =\; \bigotimes_{i=1}^N \cH_i \;  .
\ee
The operators are assumed to be of trace class, such that
\be
\label{2}
0 \; < \; {\rm Tr}_\cH |\hat A| \; < \; \infty \; .
\ee
In general, it is possible to consider different trace-class operators. However, of 
interest are only the operators containing information on particle correlations, such 
as statistical operators, reduced density operators, or correlation operators. Moreover, 
trace-class operators are bounded, hence unbounded operators are not suitable. This 
concerns, e.g., the position and momentum operators that are unbounded.   

The idea of introducing a correlation measure is based on the comparison of the global 
action of the given operator $\hat{A}$ on space (\ref{1}) and the action of its uncorrelated 
counterpart
\be
\label{3}
  \hat A^\otimes \; = \; \frac{\bigotimes_{i=1}^n \hat A_i}{({\rm Tr}_\cH \hat A)^{n-1} } 
\ee
composed of partially traced operators
\be
\label{4}
\hat A_i \; \equiv \; {\rm Tr}_{\cH\setminus\cH_i} \hat A \;   ,   
\ee
so that the normalization condition be retained,
\be
\label{5}
 {\rm Tr}_\cH \hat A^\otimes \; = \; {\rm Tr}_\cH \hat A \;  .
\ee

The correlation index for the operator $\hat{A}$ is
\be
\label{6}
\varkappa(\hat A) \; \equiv \; 
\log\; \frac{||\; \hat A\; ||}{||\; \hat A^\otimes\;||}  \; .
\ee
The norm can be taken in different forms, of which we prefer the standard operator norm
$$
 ||\; \hat A\; || \; = \; 
\sup_{\vp\in\cH} \; \frac{||\; \hat A\vp \; ||}{||\; \vp \; ||} 
\qquad (\vp \neq 0 ) \;  .
$$
In physical applications, we shall deal with Hermitian operators, for which the operator 
norm reduces to the Hermitian norm.
\be
\label{7}
 ||\; \hat A\; || \; = \; 
\sup_{\vp\in\cH} \; 
\frac{ |\; \lgl \vp\; |\; \hat A\; |\;\vp\; \rgl\;|}{\lgl\; \vp\;|\;\vp \;\rgl} 
\qquad
( \hat A^+ = \hat A) \; .
\ee
The logarithm can be taken with respect to any base. As a rule, we take the natural 
logarithm $\ln$. 

Defining correlation indices, we mainly deal with semi-positive (non-negative) operators, 
such as statistical operators, reduced density operators, and correlation operators. For
the semi-positive operators, it is easy to calculate the trace norm. Therefore the natural 
question is: Why, instead of the operator supremum norm, not to use the trace norm? Then
the definition of the correlation index would be
$$
\varkappa_1(\hat A) \; = \; \log \; \frac{||\;\hat A\;||_1}{||\hat A^\otimes\;||_1} \;  ,
$$    
where the trace norm is
$$
||\hat A\;||_1 \; \equiv \; {\rm Tr}_\cH|\;\hat A\;| \qquad
\left( |\; \hat A \; | \equiv \sqrt{\hat A^+ \; \hat A} \right) \;  .
$$
For a semi-positive operator, $|\hat{A}| = \hat{A}$, which gives
$$
 ||\hat A\;||_1 \; = \; {\rm Tr}_\cH \;\hat A\; , \qquad
||\hat A^\otimes\;||_1 \; = \; {\rm Tr}_\cH \;\hat A^\otimes 
\qquad (\hat A \geq 0 ) \;  .
$$ 
Remembering the normalization condition (\ref{5}), that is required for the idempotency 
of the projection operator,
$$
 \hat P^2_\otimes \; = \hat P_\otimes\;  \qquad
\left(  {\rm Tr}_\cH \;\hat A^\otimes = {\rm Tr}_\cH \;\hat A \right) \;  ,
$$
we come to the conclusion
$$
 \varkappa_1(\hat A) \; = \; 0 \qquad ( \hat A \geq 0 ) \;  .
$$
Thence the trace norm is not suitable for measuring correlations.

By its definition and the properties of the norm, the measure (\ref{6}) is zero for 
tensor-product operators, continuous by norm, additive, and invariant under unitary 
operations. It is semi-positive, since the transformation (\ref{3}) can be represented
as a projection
$$
 \hat A^\otimes \; = \; \hat P_\otimes \hat A \; , \qquad 
\hat P_\otimes^2 \; = \; \hat P_\otimes \;  ,
$$
where the idempotence of the projection operator can be straightforwardly checked by 
applying twice the transformation (\ref{3}). Therefore we have
$$
||\; \hat A^\otimes \; || \; = \; ||\;  \hat P_\otimes \hat A\; || \; \leq \; 
|| \; \hat P_\otimes \; || \cdot || \; \hat A\; || \; = \; ||\; \hat A \; || \;  ,
$$ 
from where the semi-positiveness of the index (\ref{6}) follows.  

Correlation indices measure all types of correlations, classical as well as quantum. Quantum 
correlations are associated with entanglement \cite{Wiliams_4,Nielsen_5,Vedral_6,Keyl_7,
Yukalov_8,Horodecki_9,Guhne_10,Wilde_11,Yukalov_12}. Below, we give several examples of 
statistical operators for quantum states, showing that the correlation indices measure 
entanglement of these quantum states. We compare the correlation indices with the von Neumann
entanglement entropy, when it is defined, that is, for binary systems with the statistical 
operator $\hat{\rho}$ acting on the Hilbert space 
$\mathcal{H} = \mathcal{H}_1 \bigotimes \mathcal{H}_2$. The von Neumann entanglement entropy 
is
\be
\label{8}
 S_E \; = \; - {\rm Tr}_{\cH_i} \hat R_i \; \ln \hat R_i \; , \qquad
\hat R_i \; \equiv \; {\rm Tr}_{\cH\setminus\cH_i} \hat\rho \;  .
\ee

For example, let us consider the {\it Einstein-Podolsky-Rosen state}
\be
\label{9}
 \hat\rho_{EPR} \; = \; \frac{1}{2} \left( \; | \; 01 \; \rgl + | \; 10\; \rgl \; \right)
\left( \; 
\lgl \; 10 \; | + \lgl \; 01\; | \; \right) \; .
\ee
Following the above procedure, we find the correlation index
\be
\label{10}
 \varkappa(\hat\rho_{EPR} ) \; = \; \ln 2 \;  .
\ee
It is worth emphasizing that this correlation index coincides with the von Neumann 
entanglement entropy equal in that case to $S_E = \ln 2$. 

For the $N$-partite {\it Bell state}, or multicat state,
\be
\label{11}
 \hat\rho_B \; = \; 
\frac{1}{2} \left( \; | \; 00\ldots 0 \; \rgl + | \; 11\ldots 1 \; \rgl \; \right)
\left( \; \lgl \; 11\ldots 1 \; | + \lgl \; 00\ldots 0 \; | \; \right) \;  ,
\ee
we get
\be
\label{12}
\varkappa(\hat\rho_B) \; = \; (N -1 ) \ln 2 \;   .
\ee
For multipartite states, the von Neumann entanglement entropy is not defined.

The {\it multipartite multimode state}, with $N$ parts and $M$ modes, 
\be
\label{13}
\hat\rho_{MM} \; = \; \frac{1}{M} \left( \sum_{n=0}^{M-1} \; | \; nn \ldots n \; \rgl\; \right)
\left( \sum_{n=0}^{M-1} \; \lgl \; nn \ldots n \; | \; \right)
\ee
leads to the correlation index
\be
\label{14}
 \varkappa(\hat\rho_{MM} ) \; =\; (N - 1) \; \ln M \;  .
\ee

Consider the $N$-particle {\it Hartree-Fock state} 
\be
\label{15}
\hat\rho_{HF} \; = \; \frac{1}{N!}
\left( \sum_{Sym} |\; 1 2\ldots N \; \rgl \right)
\left( \sum_{Sym} \lgl\; 1 2\ldots N \; | \right)
\ee
composed of the wave functions of $N$ particles in $N$ different quantum states,
$$
 \frac{1}{\sqrt{N!}} \sum_{Sym} |\; 1 2\ldots N \; \rgl \; \equiv \;
 \frac{1}{\sqrt{N!}} 
\sum_{Sym} \psi_1(x_1) \; \psi_2(x_2) \ldots \psi_N(x_N) \;  ,
$$
where the notation ``Sym" implies symmetrization (or antisymmetrization) over particles,
depending on whether they are bosons or fermions. For example, in the case of two 
particles, the wave function is
$$
\frac{1}{\sqrt{2}} \sum_{Sym} |\; 1 2\;\rgl \; = \;
\frac{1}{\sqrt{2}} \; \left[ \; \psi_1(x_1) \;\psi_2(x_2) \pm 
\psi_1(x_2) \;\psi_2(x_1) \; \right] \;  .
$$
Thus in this case the particles are in different states. This gives the 
correlation index
\be
\label{16}
\varkappa(\hat\rho_{HF} ) \; = \;  \ln \; \frac{N^N}{N!} \;    .   
\ee
For $N = 2$, this reduces to $\ln 2$, while for large $N$, it gives
\be
\label{17}
 \varkappa(\hat\rho_{HF} ) \; \simeq \; N \qquad ( N \gg 1) \; .
\ee

For the reduced Hartree-Fock state
\be
\label{18}
 \hat\rho_n \; = \; {\rm Tr}_{\cH_{n+1}} \;  {\rm Tr}_{\cH_{n+2}} \ldots
{\rm Tr}_{\cH_N} \; \hat\rho_{HF} \; ,
\ee
the correlation index is
\be
\label{19}
 \varkappa(\hat\rho_n ) \; = \; 
\ln\left[ \; \frac{(N-n)!}{N!} \; N^n \; \right] \;  ,
\ee
which for large $N$ reads as
\be
\label{20}
 \varkappa(\hat\rho_n) \; \simeq \; \frac{n(n-1)}{2N} \qquad ( N \gg 1) \;  .
\ee

It is important to notice that the correlation indices measure all types of correlations, 
quantum as well as classical. For quantum systems, they measure such quantum correlations 
as entanglement.

\section{Reduced density matrices}

Exhaustive information on the properties of statistical systems is contained in reduced
density matrices \cite{Coleman_13,Kardar_2007}. In this section, we show how correlation 
indices can be calculated on the basis of these matrices.  

The first-order reduced density matrix is defined as
\be
\label{21}
 \rho(x,x') \; =\; {\rm Tr}\; \psi(x) \; \hat\rho \; \psi^\dgr(x') \; = \;
\lgl \; \psi^\dgr(x') \; \psi(x) \; \rgl \;  ,
\ee
where $x$ is a set of variables, such as spatial coordinates, spin, etc., $\psi(x)$ is a 
field operator, the trace is over the Fock space and the angle brackets mean statistical 
averaging with a given statistical operator $\hat{\rho}$. This matrix can be treated as 
a matrix element of the reduced statistical operator
\be
\label{22}
\hat\rho_1 \; \equiv \; [ \; \rho(x,x') \; ] \; .
\ee
This operator acts on the Hilbert space 
\be
\label{23}
\cH_1 \; = \; \overline \cL \{ \; | \; k \; \rgl \; \} \; , 
\qquad
| \; k \; \rgl \; = \; [\; \vp_k(x) \; ] \; ,
\ee   
that is a closed linear envelope over a basis labeled by a multi-index $k$. It is 
straightforward to define for the space (\ref{23}) the trace,
\be
\label{24}
 {\rm Tr}_{\cH_1} \; \hat\rho_1 \; = \; 
\sum_k \; \lgl \; k \; | \; \hat\rho_1 \; | \; k \; \rgl \; = \; 
\int \rho(x,x) \; dx \; = \; N \; ,
\ee
and the norm, 
\be
\label{25}
||\; \hat\rho_1 \; || \; = \; \sup_{\vp\in\cH_1} \;
\frac{\lgl\; \vp \;|\; \hat\rho_1\;|\; \vp \;\rgl}{\lgl\; \vp \;|\; \vp \;\rgl} \; = \;
\sup_k \;
\frac{\lgl\; k \;|\; \hat\rho_1\;|\; k \;\rgl}{\lgl\; k \;|\; k \;\rgl} 
\ee
of the density operator (\ref{22}). Keeping in mind an ortho-normalized basis yields 
\be
\label{26}
||\; \hat\rho_1 \; || \; = \;  \sup_k N_k \; , 
\qquad 
N_k \; = \; \lgl \; k \; | \; \hat\rho_1 \;|\; k \;\rgl \; . 
\ee
 
The second-order reduced density matrix
\be
\label{27}
 \rho_2(x_1,x_2,x_1',x_2') \; =\; 
{\rm Tr} \;\psi(x_1) \; \psi(x_2) \; \hat\rho \; \psi^\dgr(x_2') \; \psi^\dgr(x_1') \; = \;
\lgl\; \psi^\dgr(x_2') \; \psi^\dgr(x_1') \; \psi(x_1) \; \psi(x_2) \; \rgl \;,
\ee
being considered as a matrix element, defines the second-order reduced density operator
\be
\label{28}
 \hat\rho_2 \; = \; [\; \rho_2(x_1,x_2,x_1',x_2') \; ]  
\ee
acting on the Hilbert space
\be
\label{29}
 \cH \; =\; \cH_1 \bigotimes \cH_2 \; = \; \overline\cL\{ \; |\; kp\; \rgl \; \} \;  .
\ee

Using the properties of reduced density matrices \cite{Coleman_13}, it is straightforward to 
define the trace,
\be
\label{30}
{\rm Tr}_\cH \hat\rho_2 \; = \; 
\sum_{kp} \; \lgl kp \; |\; \hat\rho_2 \;|\; kp \;\rgl \; = \;
\int \rho_2(x_1,x_2,x_1,x_2) \; dx_1 dx_2 \; = \; N ( N -1 ) \; ,
\ee
and the norm,
\be
\label{31}
||\; \hat\rho_2\; || \; = \; \sup_{kp} N_{kp} \; , 
\qquad 
N_{kp} \; \equiv \; \lgl kp \;|\; \hat\rho_2 \;|\; kp \;\rgl \;   ,   
\ee
for the second-order density operator.  

The partial traces give the operators
\be
\label{32}
 \hat R_1 \; = \; {\rm Tr}_{\cH_2} \hat\rho_2 \; = \; [ \; R_1(x_1,x_1') \;] \; ,
\qquad
 \hat R_2 \; = \; {\rm Tr}_{\cH_1} \hat\rho_2 \; = \; [ \; R_2(x_2,x_2') \;] \;  ,
\ee
whose matrix elements are
\be
\label{33}
R_1(x_1,x_1') \; = \; \int \rho_2(x_1,x_2,x_1',x_2) \; dx_2 \; = \; 
( N - 1) \; \rho(x_1,x_1') 
\ee
and
\be
\label{34}
R_2(x_2,x_2') \; = \; \int \rho_2(x_1,x_2,x_1,x_2') \; dx_1 \; = \; 
( N - 1) \; \rho(x_2,x_2') \;   .
\ee

The uncorrelated counterpart of (\ref{28}) reads as
\be
\label{35}
 \hat\rho^\otimes_2 \; = \; 
\frac{\hat R_1\bigotimes \hat R_2}{{\rm Tr}_\cH \hat\rho_2} \; = \;
\frac{N - 1}{N} \; \hat\rho_1\bigotimes \hat\rho_1 \;  .
\ee
Its trace, according to (\ref{30}), is
\be
\label{36}
{\rm Tr}_\cH \hat\rho_2^\otimes \; = \; {\rm Tr}_\cH \hat\rho_2 \; = \; N ( N - 1)
\ee
and the norm is
\be
\label{37}
||\; \hat\rho_2^\otimes \;|| \; = \; \frac{N - 1}{N} ||\; \hat\rho_1 \;||^2 \;  .
\ee

Thus, the correlation index becomes
\be
\label{38}
\varkappa(\hat\rho_2) \; = \; 
\ln \frac{||\; \hat\rho_2 \;||}{||\; \hat\rho_2^\otimes \;||} \; = \;
\ln \left[\; \left( \frac{N}{N-1} \right) \; 
\frac{||\; \hat\rho_2 \;||}{||\; \hat\rho_1 \;||^2} \; 
\right] \;   .
\ee

The value of the correlation index, certainly, depends on the physics of the system and
on the specific partition of the related Hilbert space. As an example, let us consider
the system of $N$ bosons, where each particle can be in two quantum states, so that the 
state of the system is prepared in the form of the multiparticle Bell state (\ref{11}).     
Let us treat this physical system as being composed of $N/n$ clusters of $n$ particles 
in each cluster. Since each particle can be in two states, the total number of modes of 
a single cluster of $n$ indistinguishable bosons is $n + 1$. Assume that the Bell structure 
for the system of clusters is preserved. Then the state of the whole system is described 
by the statistical operator
$$
 \hat\rho_C \; = \; \frac{1}{n+1} 
\left( \sum_{\al=0}^n |\; \al \al \ldots \al\;\rgl \right) 
\left( \sum_{\al=0}^n \lgl \; \al \al \ldots \al \;| \right) \; .
$$
The corresponding correlation index becomes
$$
\varkappa(\hat\rho_C) \; = \; \left( \frac{N}{n} - 1 \right) \; \ln (n+1) \; .
$$
This index for the clusters depends on $n$, being, generally, different from the correlation 
index (\ref{12}) for particles. If a cluster would be composed of just a single particle 
$(n = 1)$, we would come back to the correlation index (\ref{12}). If to treat the overall 
system as a single cluster with $n = N$ particles, then the correlation index is zero, which 
is evident, as far as a single cluster, representing the system, has nobody to correlate with.

\section{Systems with Bose-Einstein condensate}

In recent years, there has been high interest to systems with Bose-Einstein condensates.
Such systems, being produced in traps, allow for wide variation of their parameters. There 
has been quite a number of summarizing papers and books on the topic, of which we cite 
just several the most recent, where one can find references to the previous publications 
\cite{Dupuis_14,Yukalov_15,Yukalov_16,Castin_17,Yukalov_18}. 

In order to calculate the correlation index for a system with Bose-Einstein condensate,
we have to find the norms of the reduced density operators $\hat{\rho}_1$ and $\hat{\rho}_2$.

The energy Hamiltonian for a system of spinless atoms reads as
\be
\label{39}
\hat H \; = \; 
\int \psi^\dgr(\br) \left( -\; \frac{\nabla^2}{2m} \right) \psi(\br) \; d\br
+ \frac{1}{2} \int \psi^\dgr(\br) \;   \psi^\dgr(\br') \; \Phi(\br-\br') \;
 \psi(\br') \;  \psi(\br) \; d\br d\br' \; .   
\ee
Atomic clouds in traps usually are dilute, which allows for the representation of interactions
in the form of the local potential
\be
\label{40}
\Phi(\br) \; = \; \Phi_0 \; \dlt(\br) \; , \qquad
\Phi_0 \; = \; 4\pi \; \frac{a_s}{m} \;   ,
\ee
where $a_s$ is scattering length and $m$ is mass. Here and in what follows, the Planck constant
is set to one, $\hbar = 1$.  

The Bose-Einstein condensation is necessarily accompanied by global gauge symmetry breaking 
\cite{Lieb_19,Yukalov_20} which can, most conveniently, be realized by means of the Bogolubov 
shift \cite{Bogolubov_21,Bogolubov_22,Bogolubov_23}
\be
\label{41}
\psi(\br) \; = \; \eta(\br) + \psi_1(\br) \;   ,
\ee
in which $\eta$ is the condensate wave function and $\psi_1$ is the field operator of 
non-condensed atoms. These variables satisfy the conditions
\be
\label{42}
\lgl\; \psi_1(\br) \;\rgl \; = \; 0 \; ,
\qquad
\int\eta^*(\br) \; \psi_1(\br) \; d\br \; = \; 0 \;   .
\ee
The variables describing condensed and non-condensed atoms are normalized to the number 
of condensed atoms
\be
\label{43}
N_0 \; = \; \int |\; \eta(\br) \;|^2 \; d\br 
\ee
and the number of non-condensed atoms
\be
\label{44}
 N_1 \; = \; \lgl\; \hat N_1 \;\rgl \; = \; 
\int \lgl\; \psi_1^\dgr(\br)\; \psi_1(\br) \;\rgl \; d\br \; ,
\ee
respectively, so that the total number of atoms is
\be
\label{45}
N \; = \; N_0 + N_1 \;   .
\ee
Hence the grand Hamiltonian acquires the form
\be
\label{46}
H \; = \; \hat H - \mu_0 N_0 - \mu_1 \hat N_1 \;   .
\ee

The chemical potentials $\mu_0$ and $\mu_1$ are the Lagrange multipliers guaranteeing the 
normalizations (\ref{43}) and (\ref{44}). It is necessary to emphasize that, following Gibbs
\cite{Gibbs_24,Gibbs_25}, the grand Hamiltonian has to contain the number of Lagrange 
multipliers that is equal to the number of additional constraints imposed on the system. 
Only then the system description is self-consistent and uniquely defined, which is assured 
by the Shore-Johnson theorem \cite{Shore_26}. Such statistical ensembles that uniquely define 
the system are called representative \cite{Ter_27,Yukalov_28}. If the number of Lagrange 
multipliers is less then the number of variables, the description cannot be self-consistent 
and uniquely defined, as has been demonstrated for Bose-condensed systems by the Hohenberg-Martin 
dilemma \cite{Hohenberg_29} showing that a Bose-condensed system with a single Lagrange 
multiplier is either unstable, when the number of condensed atoms $N_0$ does not minimize 
the thermodynamic potential, or the condensate cannot exist, when the Hugenholtz-Pines 
relation \cite{Hugenholtz_30} does not hold. For a system with Bose-Einstein condensate, 
the global gauge symmetry is necessarily broken \cite{Yukalov_18,Lieb_19,Yukalov_20}, which, 
after the Bogolubov shift, results in two variables, the condensate wave function and the 
operator of non-condensed atoms. The condensate function is normalized to the number of 
condensed atoms $N_0$ that has to minimize the thermodynamic potential, which is a condition 
of the system stability, according to the Bogolubov-Ginibre theorem 
\cite{Bogolubov_21,Bogolubov_22,Ginibre_31}. Gauge symmetry breaking leads to the appearance 
of  the Bose-Einstein condensate and, due to the given total number of atoms $N$, to the 
normalization (\ref{44}) for the number of non-condensed atoms $N_1 = N - N_0$. The existence 
of the Bose-Einstein condensate, hence of gauge symmetry breaking, leads to the 
Hugehnholtz-Pines relation \cite{Hugenholtz_30}. Taking account of all required constraints
results in a self-consistent approach  
\cite{Yukalov_18,Yukalov_28,Yukalov_32,Yukalov_33,Yukalov_34,Yukalov_35} to be used below for
calculating the correlation index of a Bose-condensed system.

The first-order density operator 
\be
\label{47}
\hat\rho_1 \; = \; [\; \rho(\br,\br') \;]
\ee
enjoys the matrix elements
\be
\label{48}
 \rho(\br,\br') \; = \; \eta^*(\br') \; \eta(\br) + \rho_1(\br,\br') \;  ,
\ee
in which 
\be
\label{49}
\rho_1(\br,\br') \; \equiv \; \lgl\; \psi_1^\dgr(\br') \; \psi_1(\br) \;\rgl \;   .
\ee
Because of the broken gauge symmetry, there also appears the anomalous average
\be
\label{50}
\sgm_1(\br,\br') \; \equiv \; \lgl\; \psi_1(\br') \; \psi_1(\br) \;\rgl \;   .
\ee

Nowadays, ultracold atomic gases can be trapped in box-shaped optical potentials, inside
which the gas represents a finite uniform system of $N$ atoms in volume $V$
\cite{Meyrath_36,Van_37,Gaunt_38,Chomaz_39,Navon_40,Mukherjee_41,Hueck_42,Tajik_43,Bause_44,
Navon_45,Gauthier_46}. For a uniform gas, the condensate function is a constant
\be
\label{51}
 \eta(\br) \; = \; \sqrt{\rho_0} \; , \qquad \rho_0 \; \equiv \; \frac{N_0}{V} \;  ,  
\ee
and the field operators of non-condensed atoms can be expanded over plane waves,
\be
\label{52}
\psi_1(\br) \; = \; \sum_{k\neq 0} a_k \vp_k(\br) \; ,
\qquad
\vp_k(\br) \; = \; \frac{1}{\sqrt{V}} \; e^{i\bk\cdot\br} \;   ,
\ee
giving
\be
\label{53}
 \rho_1(\br,\br') \; =\; \sum_{k\neq 0} n_k \; \vp_k(\br) \; \vp_k^*(\br') \; ,
\qquad
 \sgm_1(\br,\br') \; =\; \sum_{k\neq 0} \sgm_k \; \vp_k(\br) \; \vp_k(\br') \; .
\ee
Employing the Hartree-Fock-Bogolubov (HFB) approximation leads to the normal and anomalous 
averages
$$
n_k \; = \;  \lgl\; a_k^\dgr \; a_k \;\rgl \; = \;
\frac{\om_k}{2\ep_k}\; \coth\left( \frac{\ep_k}{2T}\right) - \; \frac{1}{2} \; ,
$$
\be
\label{54}
 \sgm_k \; = \;  \lgl\; a_{-k} \; a_k \;\rgl \; = \; -\;
\frac{mc^2}{2\ep_k}\; \coth\left( \frac{\ep_k}{2T}\right) \;  ,
\ee
where the notation is used
\be
\label{55}
\om_k \; =\; mc^2 + \frac{k^2}{2m}
\ee
and the spectrum of collective excitations is
\be
\label{56}
\ep_k \; = \; \sqrt{ (ck)^2 + \left( \frac{k^2}{2m} \right)^2} \;   .
\ee
The sound velocity $c$ is defined by the equation
\be
\label{57}
mc^2 \; =\; \Phi_0 \; ( \rho_0 + \sgm_1 ) \;   ,
\ee
with the densities
$$
\rho_0 \; = \; \rho - \rho_1 \; , 
\qquad 
\rho \; \equiv \; \frac{N}{V} \; 
\qquad \rho_1 \; \equiv \; \frac{N_1}{V} \; ,
$$
\be
\label{58}
\rho_1 \; = \; \frac{1}{V} \sum_{k\neq 0} n_k \; , 
\qquad
\sgm_1 \; = \; \frac{1}{V} \sum_{k\neq 0} \sgm_k \;  .
\ee

Let us consider, for simplicity, the zero-temperature case. Then
\be
\label{59}
 n_k \; =\; \frac{\om_k-\ep_k}{2\ep_k} \; , 
\qquad 
\sgm_k \; = \; - \; \frac{mc^2}{2\ep_k} \qquad (T = 0)  
\ee
and 
\be
\label{60}
\rho_1 \; = \; \frac{(mc)^3}{3\pi^2} \; , 
\qquad 
\sgm_1 \; = \; - mc^2 \int \frac{1}{2\ep_k} \; \frac{d\bk}{(2\pi)^3} \; .
\ee
  
It is convenient to introduce the dimensionless fractions
\be
\label{61}
n_0 \; \equiv \; \frac{\rho_0}{\rho} \; , \qquad  
n_1 \; \equiv \; \frac{\rho_1}{\rho} \; , \qquad 
\sgm \; \equiv \; \frac{\sgm_1}{\rho} \; ,
\ee
dimensionless sound velocity
\be
\label{62}
s \; \equiv \; \frac{mc}{\rho^{1/3}} \;   ,
\ee
and the dimensionless coupling parameter
\be
\label{63}
 g \; \equiv \; \rho^{1/3} \; a_s \;  .
\ee
The fractions of condensed and non-condensed atoms become
\be
\label{64}
 n_0 \; = \; 1 - \; \frac{s^3}{3\pi^2} \; , \qquad 
n_1 \; \equiv \; \frac{s^3}{3\pi^2} \;  .
\ee
The integral in (\ref{60}), defining the anomalous average $\sigma_1$ is divergent, but can 
be straightforwardly regularized by resorting to dimensional regularization \cite{Yukalov_35},
which results in the expression
\be
\label{65}
 \sgm \; = \; \frac{8}{\sqrt{\pi} } \; g^{3/2} \; \left( n_0 +
\frac{8}{\sqrt{\pi}}\; g^{3/2} \; \sqrt{n_0} \right)^{1/2} \;  .
\ee
For the dimensionless sound velocity, we get
\be
\label{66}
 s^2 \; = \; 4\pi g( n_0 + \sgm) \;  .
\ee

The single-atom occupation number takes the form
\be
\label{67}
 N_k \; = \; \lgl\; k \;|\; \hat\rho_1 \;|\; k \;\rgl \; = \;
N_0 \dlt_{k0} + n_k \;  .
\ee
From expressions (\ref{59}), we have
\be
\label{68}
\sup_k n_k \; = \; \sup_k |\; \sgm_k \;| \; = \; \frac{mc}{2k_{min}} \; , 
\ee
with the minimal wave vector 
\be
\label{69}
 k_{min} \; = \; \frac{2\pi}{V^{1/3} } \; = \; 
2\pi \left( \frac{\rho}{N} \right)^{1/3} \;  .
\ee
Therefore
\be
\label{70}
\sup_k n_k \; = \; 
\sup_k |\; \sgm_k \;| \; = \; \frac{s}{4\pi} \; N^{1/3} \;   .
\ee

In this way, the norm of the first-order reduced density operator is
\be
\label{71}
 ||\; \hat\rho_1 \;|| \; = \; 
\sup\left\{ n_0 N; ~ \frac{s}{4\pi}\; N^{1/3} \right\} \;  .
\ee

The second-order reduced density matrix, after the Bogolubov shift and the HFB approximation,
reads as
$$
\rho_2(\br_1,\br_2,\br_1',\br_2') \; = \;
$$
$$
= \; 
\rho_0^2 + \rho_0 \; \left[ \; \rho_1(\br_1,\br_1') +
\rho_1(\br_1,\br_2') + \rho_1(\br_2,\br_1') + \rho_1(\br_2,\br_2') +
\sgm_1(\br_1,\br_2) + \sgm_1^*(\br_1',\br_2') \; \right] +
$$
\be
\label{72}
+ 
\rho_1(\br_1,\br_1')\; \rho_1(\br_2,\br_2') + \rho_1(\br_1,\br_2')\; \rho_1(\br_2,\br_1')
+ \sgm_1(\br_1,\br_2)\; \sgm_1^*(\br_1',\br_2') \; .
\ee
The related occupation number is
\be
\label{73}
N_{kp} \; = \; \lgl\; kp \;|\; \hat\rho_2 \;|\; kp \;\rgl \; = \;
N_0^2 \dlt_{k0} \; \dlt_{p0} + N_0 \; ( n_k \dlt_{p0} + n_p \dlt_{k0} ) +
n_k n_p + \left( n_k^2 + |\; \sgm_k \;|^2 \right) \; \dlt_{kp} \;   .
\ee
Then we get the norm
\be
\label{74}
||\; \hat\rho_2 \;|| \; = \; \sup\left\{ n_0^2 \; N^2 \; ; ~
\frac{n_0s}{4\pi} \; N^{4/3} \; ; ~ \frac{3s^2}{16\pi^2} \; N^{2/3} \right\} \;  .
\ee

Thus for the correlation index (\ref{38}) we obtain
\be
\label{75}
 \varkappa(\hat\rho_2) \; = \;
\ln \left[\; \left( \frac{N}{N-1} \right) \;
\frac{\sup\{n_0^2N^2;~ \frac{n_0s}{4\pi}\; N^{4/3}; ~ \frac{3s^2}{16\pi^2}\; N^{2/3} \} }
{ \sup\{ n_0^2 N^2; ~ \frac{s^2}{16\pi^2}\; N^{2/3} \}  } \; \right] \; .
\ee
The condensate fraction $n_0$ and sound velocity $s$ are defined by equations (\ref{64}),
(\ref{65}), and (\ref{66}), so that the correlation index is a function of the number of 
atoms $N$ and the coupling parameter $g$.

The behavior of the correlation index (\ref{75}), as a function of the number of particles
$N$, at different coupling parameters $g$, is shown in Fig. 1. The correlation index 
decreases with the growing number of atoms $N$. This happens because, with increasing $N$, 
long-range interactions become prevalent over local (short-range) interactions, as a result 
of which correlation functions decouple into products, thus diminishing the correlation 
index. 

\begin{figure}[ht]
\centerline{
\includegraphics[width=11cm]{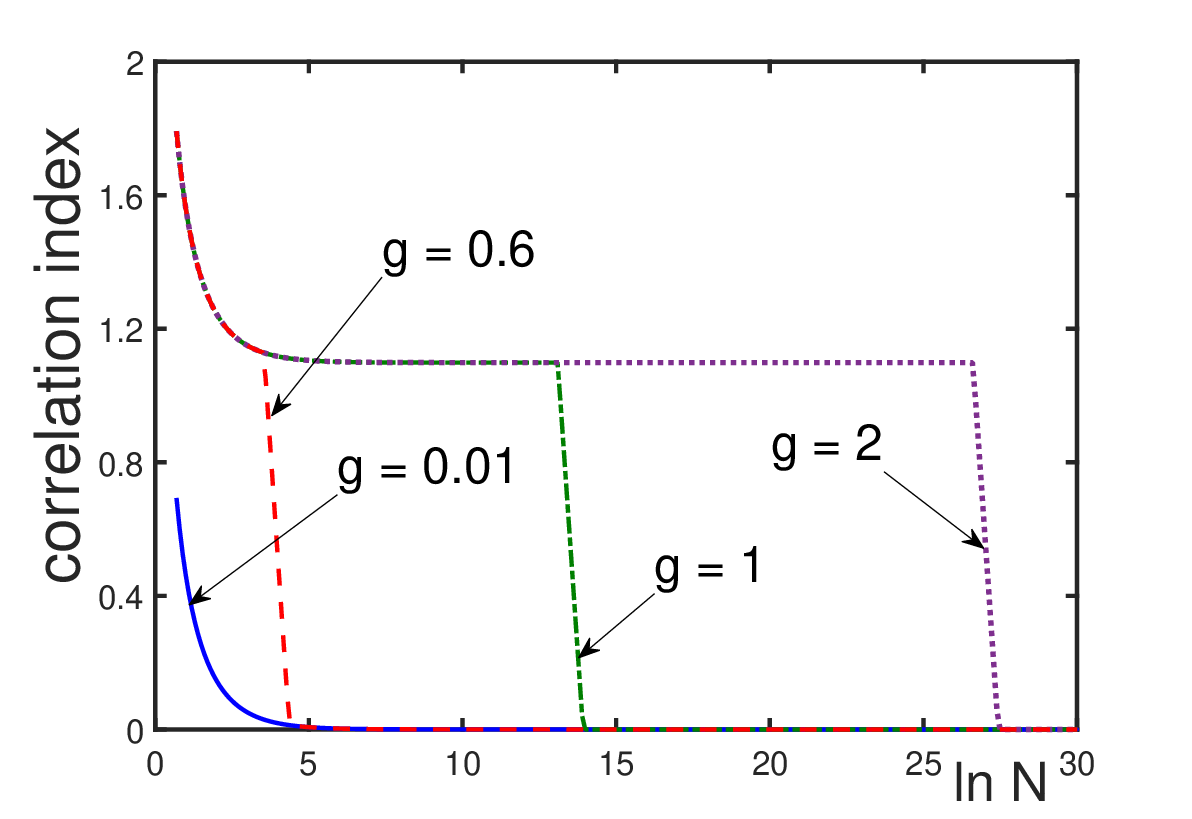} }
\caption{\small
Behavior of the correlation index $\varkappa(\hat\rho_2)$ as a function of $\ln N$ for different 
coupling parameters $g$.
}
\label{fig:Fig.1}
\end{figure}

The correlation index, as a function of the coupling parameter $g$, at different number of 
atoms $N$, is presented in Fig. 2. It increases with the stronger coupling parameter $g$. 
This increase is due to the suppression of the condensate fraction with growing $g$, so 
that the terms in the correlation index containing $n_0$ diminish as compared to the term 
describing local interactions.  
  
\begin{figure}[ht]
\centerline{
\includegraphics[width=11cm]{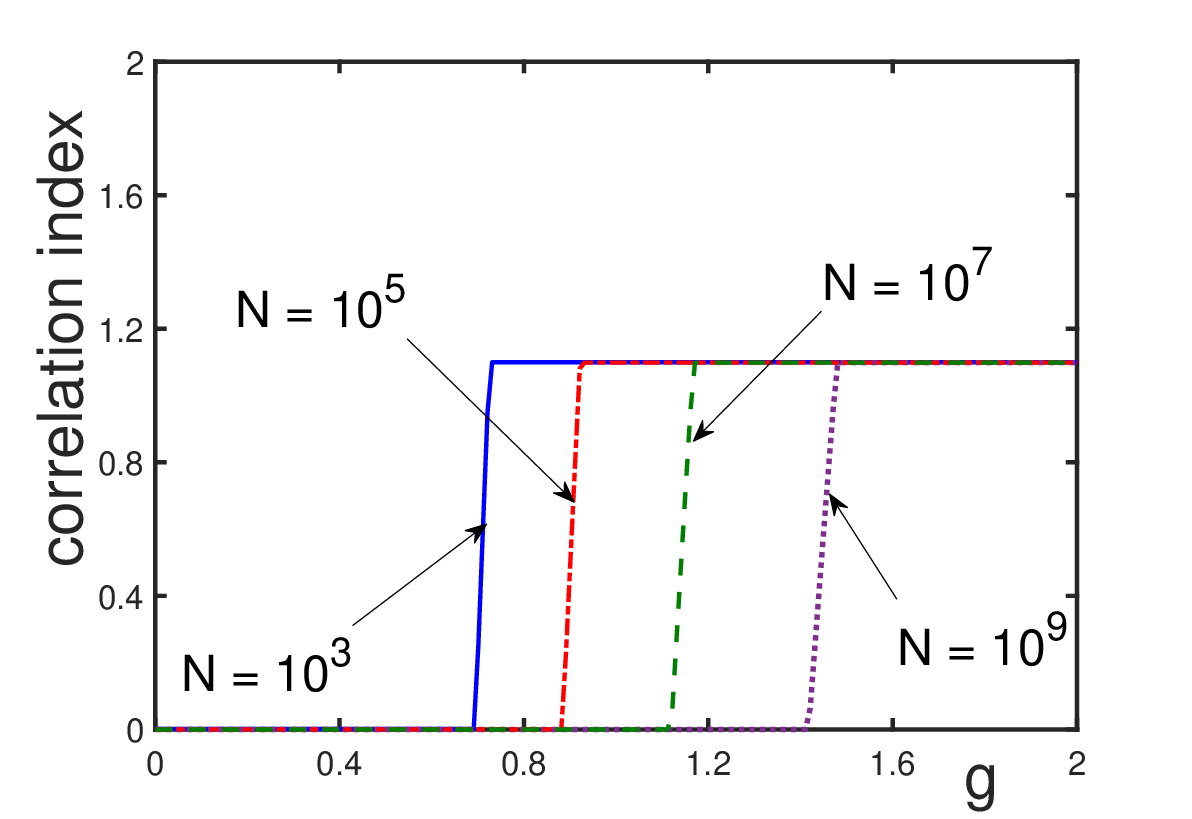} }
\caption{\small
Behavior of the correlation index $\varkappa(\hat\rho_2)$ as a function of the coupling 
parameter $g$ for different $N$.
}
\label{fig:Fig.2}
\end{figure}

The behavior of the correlation index can be employed for characterising the systems of 
trapped Bose-condensed atoms, for which both parameters, the number of trapped atoms, $N$, 
and the interaction strength, $g$, can be varied in a wide range.

\section{Superconducting systems}

Calculating the norms of the reduced density matrices, when there appear anomalous 
averages, it is important to accurately take into account pair particle correlations,
since their occurrence is, sometimes, crucial for the value of the correlation indices. 
From the technical point of view, this is related to the choice of appropriate basis
of wave functions. As an illustration, let us consider the case of a superconducting 
system of $N$ particles with spin $1/2$, keeping in mind the Bardeen-Cooper-Schrieffer 
Hamiltonian.    

To find the correlation index, we need to start with a second-order reduced density
operator
\be
\label{S1}
 \hat\rho_2 \; = \; [\; \rho_2(\br_1,\br_2,\br_1',\br_2') \; ] \;  ,
\ee
whose matrix elements are
\be
\label{S2}
 \rho_2(\br_1,\br_2,\br_1',\br_2') \; = \; \lgl\; \psi_{-s}^\dgr(\br_2') \;
 \psi_{s}^\dgr(\br_1') \; \psi_{s}(\br_1)\; \psi_{-s}(\br_2) \;\rgl \; ,
\ee 
where the index $s$ labels the particle spin.

In the Hartree-Fock-Bogolubov approximation, we have
\be
\label{S3}
\rho_2(\br_1,\br_2,\br_1',\br_2') \; = \; \rho(\br_1,\br_1') \; \rho(\br_2,\br_2') \; + \;
\sgm^*(\br_2',\br_1')\; \sgm(\br_2,\br_1)\; ,
\ee
with the normal average
\be
\label{S4}
\rho(\br,\br') \; = \; 
\lgl\; \psi_{s}^\dgr(\br') \; \psi_{s}(\br) \; \rgl \; = \;
\frac{1}{V} \sum_k n_k \; e^{i\bk\cdot(\br-\br')}
\ee
and the anomalous average
\be
\label{S5}
\sgm(\br,\br') \; = \; 
\lgl\; \psi_{-s}(\br') \; \psi_{s}(\br) \; \rgl \; = \;
\frac{1}{V} \sum_k \sgm_k \; e^{i\bk\cdot(\br-\br')} \;  .
\ee

The trace of the correlation operator (\ref{S1}) reads as
\be
\label{S6}
{\rm Tr}_\cH \hat\rho_2 \; = \; 
\int \rho_2(\br_1,\br_2,\br_1,\br_2) \; d\br_1 d\br_2  \; = \;
N^2 + \frac{1}{2} \; n_\pi N \; ,
\ee   
where the notation is used for the fraction of pair-correlated particles
\be
\label{S7}
n_\pi \; \equiv \; \frac{2N_\pi}{N}
\ee
and for the total number of the pair-correlated particles
\be
\label{S8}
N_\pi \; = \; \int |\; \sgm(\br_2,\br_1) \; |^2 d\br_1 d\br_2 \; = \;
 \sum_k |\; \sgm_k \; |^2 .
\ee

To find the norm of the second-order operator (\ref{S1}), one has to solve the related
eigenproblem \cite{Yukalov_2025}, which gives the eigenfunction
\be
\label{S9}
 \vp(\br_1,\br_2) \; = \; \frac{\sgm(\br_2,\br_1)}{\sqrt{N_\pi} } \;  .
\ee
For the norm, we obtain
\be
\label{S10}
||\; \hat\rho_2 \; || \; = \; \frac{1}{2} \; n_\pi N + \frac{1}{4} \; .
\ee
Note that if one would use as an approximate eigenfunction the structure formed by plane 
waves, as for independent particles, one would lose from (\ref{S10}) the most important 
term proportional to $N$. 

The partially traced density matrices are
$$
 R(\br_1,\br_1') \; = \; 
\int \rho_2(\br_1,\br_2,\br_1',\br_2)\; d\br_2 \; = \; 
N \; \rho(\br_1,\br_1') + 
\int \sgm^*(\br_2,\br_1')\; \sgm(\br_2,\br_1)\; d\br_2 \; = \;
$$
\be
\label{S11}
\; = \; 
\frac{1}{V} 
\sum_k \left( N n_k + |\; \sgm_k \; |^2 \right) e^{i\bk\cdot(\br_1-\br_1')} 
\ee
and, similarly derived, 
\be
\label{S12}
  R(\br_2,\br_2') \; = \; 
\int \rho_2(\br_1,\br_2,\br_1,\br_2')\; d\br_1 \; = \; 
\frac{1}{V} 
\sum_k \left( N n_k + |\; \sgm_k \; |^2 \right) e^{i\bk\cdot(\br_2-\br_2')} \; .
\ee
Using these expressions as matrix elements yields the partial operators
\be
\label{S13}
\hat R_1 \; = \; [\;  R(\br_1,\br_1') \;] \; , \qquad
\hat R_2 \; = \; [\;  R(\br_2,\br_2') \;] \;  ,
\ee
whose norms are
\be
\label{S14}
||\; \hat R_i \; ||  \; = \; 
\sup_k \lgl \; k \;|\; \hat R_i \;|\; k \; \rgl \; = \;
\sup_k \left( N n_k + |\; \sgm_k \; |^2 \right) \; .
\ee
For the norm of the uncorrelated second-order density operator we get
\be
\label{S15}
||\; \hat\rho_2^\otimes \; || \; = \; 
\frac{||\; \hat R_i \; ||^2}{{\rm Tr}_\cH\hat\rho_2} \; = \; 
\sup_k \frac{(N n_k + |\; \sgm_k \; |^2)^2}{N^2+ \frac{1}{2} \; n_\pi N} \; .
\ee 
 
In this way, for the correlation index we find
\be
\label{S16}
\varkappa(\hat\rho_2) \; = \; \ln
\frac{(\frac{1}{2} \; n_\pi N + \frac{1}{4}) ( N^2 + \frac{1}{2} \; n_\pi N)}
{\sup_k(N n_k + |\; \sgm_k \; |^2)^2} \;  .
\ee
At low temperature, one approximately has
$$
\sup_k( N n_k + |\; \sgm_k \; |^2 ) \cong  \frac{1}{2} \; N + \frac{1}{4} \; .
$$
Therefore the correlation index becomes
\be
\label{S17}
\varkappa(\hat\rho_2) \; = \; 
\ln \frac{2N(2n_\pi N + 1)(2N + n_\pi)}{(2N+1)^2} \;   .
\ee

The absence of the correlated pairs implies non-superconducting state, when the 
correlation index is close to zero,
\be
\label{S18}
\varkappa(\hat\rho_2) \; \simeq \; 0 \qquad ( n_\pi = 0, ~ N \ra\infty) \; .
\ee
But for the superconducting state, where there exist pair-correlated particles, the 
correlation is large, 
\be
\label{S19}
\varkappa(\hat\rho_2) \; \simeq \; \ln (2n_\pi N) \qquad
( n_\pi > 0, ~ N \ra\infty) \;  ,
\ee
behaving as $\ln N_\pi$. The physical meaning of this behavior seems rather clear--with 
the increasing number of particles, the number of pair-correlated particles $N_\pi$ grows, 
hence the level of correlations, as compared to the system of uncorrelated particles, also 
increases.

\section{Correlations in spin systems}

Correlation indices can also be introduced for spin systems. As an illustration, let us 
consider particles in a lattice, with spin one-half characterized by the spin operators 
$S_i^z$, with $i = 1,2,\ldots$ enumerating the lattice vectors $\bf{a}_i$. A first-order 
correlation operator can be defined as
\be
\label{76}
 \hat C_1 \; =\; [\; C_{ij} \; ] \; , \qquad
C_{ij} \; =\; \lgl\; S_i^z \; S_j^z \;\rgl \;  ,
\ee
where, as usual, the angle brackets imply statistical averaging. This operator acts on 
the Hilbert space
\be
\label{77}
\cH_1 \; =\; \overline\cL\{ \; |\; k \;\rgl \; \}   
\ee
that is a closed linear envelope over the basis functions
\be
\label{78}
 |\; k \;\rgl \; = \; [\; \vp_k(\ba_i) \; ] \; , \qquad
\vp_k(\ba_i) \; = \; \frac{1}{\sqrt{N}} \; e^{i\bk \cdot \ba_i} \;  .
\ee
The average reduced spin variable is denoted as
\be
\label{79}
 z \; \equiv \; \frac{2}{N} \sum_{i=1}^N \; \lgl\; S_i^z \;\rgl \; .
\ee

In order not to be limited by the choice of the lattice type, let us treat the system in
the mean-field approximation, keeping in mind long-range spin interactions. Then we get the 
matrix elements 
\be
\label{80}
\lgl\; k \;|\; \hat C_1 \;|\; p \;\rgl \; = \; 
\frac{1}{4} \;\left[\; N z^2 \dlt_{k0} \dlt_{p0} +
(1 - z^2 ) \; \dlt_{pk} \; \right] \;   .
\ee
Therefore the norm and trace of (\ref{76}) are
\be
\label{81}
 ||\; \hat C_1 \; || \; = \; \frac{1}{4} \; \left[ \; ( N -1 ) z^2 + 1 \; \right] \; ,
\qquad
{\rm Tr}_{\cH_1} \hat C_1 \;  = \; \frac{N}{4} \;  .
\ee

The second-order correlation operator  
\be
\label{82}
 \hat C_2 \; = \; [ \; C_{ijmn} \;] \; , \qquad
C_{ijmn} \; = \; \lgl\; S_i^z \; S_j^z \; S_m^z \; S_n^z \; \rgl  
\ee
acts on the Hilbert space $\mathcal{H}$ defined as in (\ref{29}). This gives the norm 
and trace
\be
\label{83}
 ||\; \hat C_2 \; || \; = \; \frac{1}{16} \; \left( N^2 z^4 + 6 Nz^2 + 6 \right) \; ,
\qquad
{\rm Tr}_\cH \hat C_2 \;  = \; \frac{N^2}{16} \;  .
\ee

The uncorrelated operator (\ref{3}) takes the form
\be
\label{84}
 \hat C_2^\otimes \; = \; \hat C_1 \bigotimes \hat C_1 \;  ,
\ee
which gives the norm
\be
\label{85}
||\; \hat C_2^\otimes \;|| \; = \; ||\; \hat C_1 \;||^2 \; = \;
\frac{1}{16} \; ( Nz^2 + 1 )^2 \;   .
\ee

In this way, we find the correlation index
\be
\label{86}
\varkappa(\hat C_2 ) \; = \;
 \ln \; \frac{||\; \hat C_2\;||}{||\; \hat C_2^\otimes\;||} \; = \; 
\ln \; \frac{N^2 z^4 + 6 N z^2 +6}{N^2 z^4 + 2 N z^2 + 1} \; .
\ee
The index is maximal in the nonmagnetic phase, when $z \ra 0$ and 
$\varkappa(\hat{C}_2) \ra \ln 6$. For the ordered phase, when $N z^2 \ra \infty$, the 
correlation index tends to zero. This happens since in that limit the mean-field
approximation, with long-range forces, becomes exact and correlation functions decouple
making the correlation operator $\hat{C}_2$ close to the uncorrelated operator (\ref{84}). 
The correlation index (\ref{86}) monotonically decreases as a function of $Nz^2$, which 
is seen from expression (\ref{86}).

\section{Correlations in nonequilibrium systems}

It is useful to emphasize that correlation indices can be introduced for nonequilibrium 
as well as for equilibrium systems. As examples, let us consider nonlinear dynamics from 
a strongly nonequilibrium metastable state that is demonstrated by several physical systems,
whose dynamics enjoy common features. We keep in mind the following setups. First, we can 
consider a magnetic cluster inserted into a magnetic coil connected with a resonance electric 
circuit, as is described in detail in Refs. \cite{Yukalov_47,Yukalov_48}. Second, a similar 
setup can be realized for a ferroelectric cluster in a resonant cavity \cite{Yukalov_49}.
Finally, the relaxation of two-level atoms and quantum dots from a nonequilibrium metastable 
state in the Dicke setup can also be represented as the motion of pseudospin variables 
\cite{Yukalov_50,Yukalov_51}. In all these setups the problem reduces to the study of the 
nonlinear dynamics for the pseudospin variables $z$, defined in (\ref{79}), and for the 
variable
\be
\label{87}
 w  \; \equiv \;  \frac{4}{N^2} \sum_{i\neq j}^N \; \lgl\; S_i^+ \; S_j^- \;\rgl \; ,
\ee
where $S_i^\pm$ are the ladder operators. 
  
To this end, in addition to the correlation index (\ref{86}), we need to investigate the
correlation index associated with the ladder operators. For this purpose, let us consider
the correlation operator 
\be
\label{88}
 \hat W_1 \; = \; [\; W_{ij} \;] \; , \qquad 
W_{ij} \; =\; \lgl\; S_i^+ \; S_j^- \;\rgl \;  .
\ee
Its norm and trace, respectively, are
\be
\label{89}
||\; \hat W_1 \; || \; =\; \frac{1}{2} ( 1 + z) + \frac{N}{4} \; w \; ,
\qquad
{\rm Tr}_{\cH_1} \hat W_1 \; =\; \frac{N}{2} \; ( 1 + z) \;   .
\ee

Also, we need to consider the second-order correlation operator
\be
\label{90}
\hat W_2 \; =\; [ \; W_{ijmn} \; ] \; , \qquad
W_{ijmn} \; =\; \lgl\; S_i^+ \; S_j^+ \; S_m^- \; S_n^- \;\rgl \;   ,
\ee
which has the norm
\be
\label{91}
||\; \hat W_2 \; || \; =\; \frac{N^2}{16}\; w^2 + \frac{N}{4} \;( 1 + z) \; w +
\frac{1}{2} \; ( 1 + z)^2 
\ee
and the trace
\be
\label{92}
 {\rm Tr}_\cH \hat W_2 \; = \; 
\sum_{ij} \; \lgl  S_i^+ \; S_j^+ \; S_j^- \; S_i^- \;\rgl \; = \;
\frac{N^2}{4} \; ( 1 + z)^2  \;  .
\ee

For the uncorrelated operator (\ref{3}), we find
\be
\label{93}
 \hat W_2^\otimes \; =\; \hat W_1 \bigotimes \hat W_1 \;  .
\ee
The related norm is
\be
\label{94}
||\; \hat W_2^\otimes \;|| \; = \;  ||\; \hat W_1 \;||^2 \; = \;
\left[ \; \frac{N}{4} \; w + \frac{1}{2} \; ( 1 + z) \; \right]^2 \;  .
\ee

Thus, for the correlation index, we get
\be
\label{95}
 \varkappa(\hat W_2) \; = \;
\ln \; \frac{N^2 w^2 + 4 N(1+z) w + 8(1+z)^2}{[\; Nw + 2( 1+z)\;]^2} \;  .
\ee
The index is maximal, reaching $\ln 2$, when $w \ra 0$, and it is zero, when $N w \ra \infty$.

To study the temporal behavior of both indices (\ref{86}) and (\ref{95}), we need to know 
the dynamics of the variables $z$ and $w$. From the detailed analysis of the related 
equations of motion it follows \cite{Yukalov_47,Yukalov_48,Yukalov_49,Yukalov_50,Yukalov_51}
that the typical temporal behavior of these variables can be approximated as
\be
\label{96}
 z(t) \; = \; - \tanh\left( \frac{t-t_0}{\tau_0} \right) \;  ,
\qquad 
w(t) \; = \; {\rm sech}^2\left( \frac{t-t_0}{\tau_0} \right) \; , 
\ee
with the parameters $t_0$ and $\tau_0$ depending on the type of the studied system. For the 
mentioned above setups, measuring time in units of $\tau_0$, the dimensionless temporal 
behavior of $z$ and $w$ can be approximated by the functions    
\be
\label{97}
z \; = \; - \tanh(t-1) \; , \qquad w \; = \; {\rm sech}^2(t-1) \;  .
\ee
The temporal behavior of the correlation indices (\ref{86}) and (\ref{95}), as functions 
of time measured in units of $\tau_0$, is presented in Figs. 3 and 4, respectively. The 
correlation index (\ref{86}) is maximal when the variable $z$ crosses zero, while the 
correlation index (\ref{95}) monotonically diminishes with time, being maximal at the
beginning of the process.    

\begin{figure}[ht]
\centerline{
\includegraphics[width=11cm]{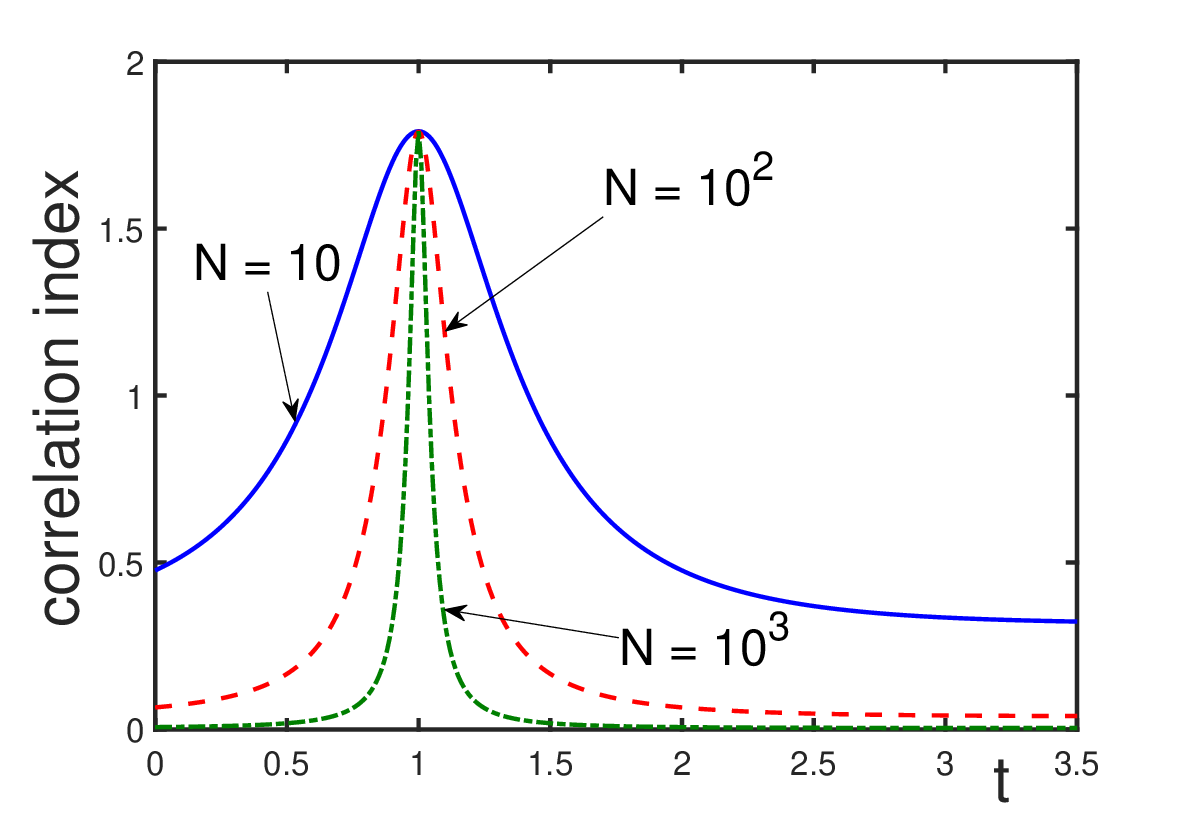} }
\caption{\small
Behavior of the correlation index $\varkappa(\hat C_2)$ as a function of dimensionless time $t$ 
for different $N$.
}
\label{fig:Fig.3}
\end{figure}   

\begin{figure}[ht!]
\centerline{
\includegraphics[width=11cm]{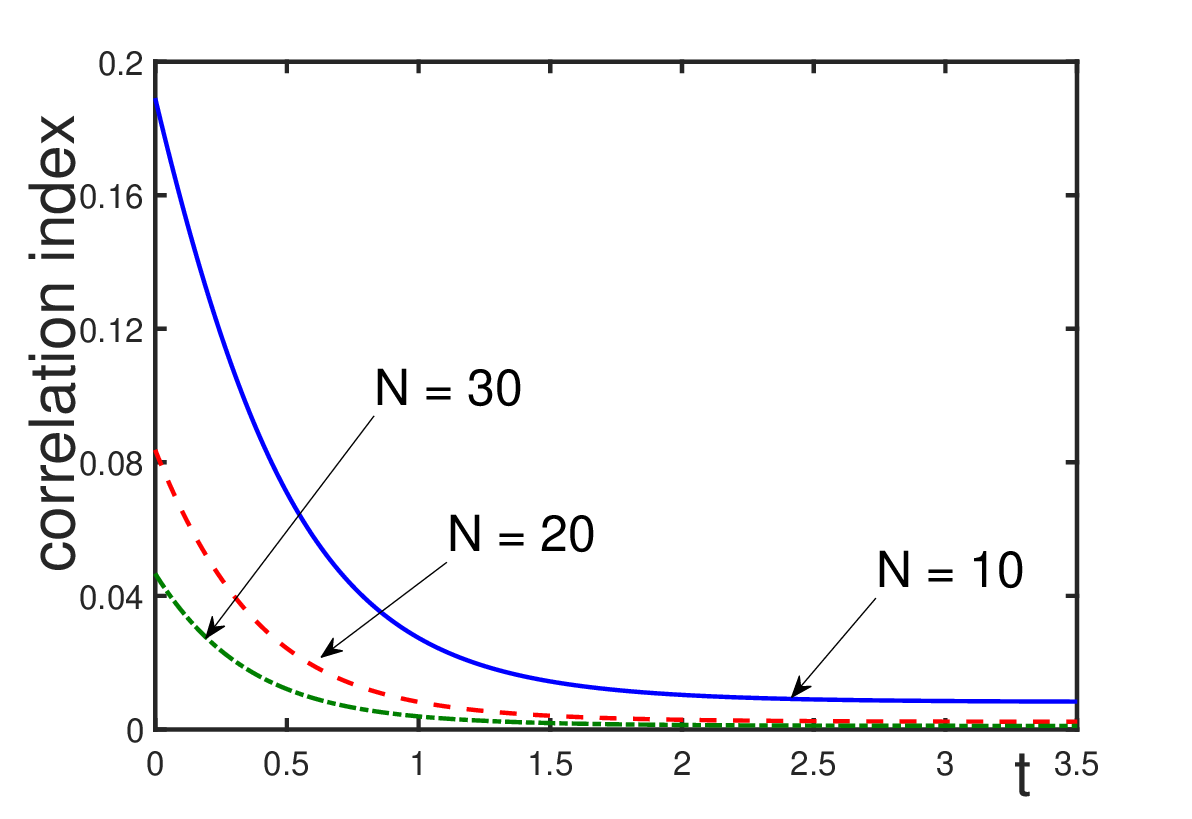} }
\caption{\small
Behavior of the correlation index $\varkappa(\hat W_2)$ as a function of dimensionless time $t$ 
for different $N$.
}
\label{fig:Fig.4}
\end{figure}

\section{Conclusion}

The notion of correlation indices is introduced, which serves as a measure of the correlation 
strength between the parts of a composite system. The main feature of the notion of the 
correlation indices is their generality. They measure all types of correlations, quantum 
as well as classical, being defined for quantum as well as for classical systems, for 
bipartite as well as for multipartite systems, for equilibrium as well as for nonequilibrium 
processes. In the domain of quantum systems they include the measurement of quantum 
entanglement of multipartite systems. Recall that the von Neumann entropy measures 
entanglement only for bipartite systems. Similarly, the partially traced quadratic Renyi 
entropy also measures entanglement only for bipartite quantum systems.

The calculation of correlation indices is demonstrated for a nontrivial case of trapped Bose
atoms with Bose-Einstein condensate and for a superconducting system. The temporal behavior 
of correlation indices for nonequilibrium systems is exemplified by nonlinear dynamics of 
spin or pseudospin variables. 

The correlation indices provide a measure unambiguously characterizing correlation strength
in different systems, whether finite or in the thermodynamic limit. Finite systems can be 
considered in two setups, in free space and in traps. Atomic systems confined in traps
nowadays are intensively studied providing rich possibility of engineering the system 
properties, including particle interactions, density, temperature, and boundary conditions.
Traps can have different shapes. There are box-shape traps imitating with high precision
uniform systems, although containing a finite number of atoms. In trapped systems, the 
boundary conditions are regulated by trapping fields, so that specific surface effects,
existing in finite systems in free space, can be suppressed. For finite systems in free
space, surface effects can also be neglected when the number of particles is large ($N \gg 1$).
In any case, the role of surface effects, even if they occur, does not change the basic 
behavior of the correlation indices.       

As interesting future applications of the correlation indices, we can mention the following 
possibilities. Finite systems of trapped bosonic atoms at low temperature, the related density 
matrices and their properties are widely studied in numerical calculations using the 
multiconfigurational time-dependent Hartree approach \cite{Alon_52,Sakman_53,Lode_54,Alon_55}. 
It seems that, having in hands reduced density matrices and their eigenvalues, it would be
straightforward to calculate the correlation indices for the interesting static and dynamic
setups considered for bosonic atoms.   
   
In papers \cite{Alon_52,Sakman_53,Lode_54,Alon_55}, the authors describe a numerical method 
of calculating the wave functions, reduced density matrices and their eigenvalues for 
trapped atoms. Knowing the eigenvalues $N_k$ and $N_{kp}$ of the first-order and second-order 
density matrices, the correlation index is calculated by the expression 
$$
\varkappa(\hat\rho_2) \; = \; 
\ln \; \frac{||\;\hat\rho_2\;||}{||\;\hat\rho_2^\otimes\;||} \; = \;
\ln \left[\; \left(\frac{N}{N-1}\right) \;
\frac{||\; \hat\rho_2\;||}{\; ||\hat\rho_1\;||^2} \;\right] \; = \;
\ln \left[\;  \left(\frac{N}{N-1}\right) \; \frac{\sup_{kp}N_{kp}}{(\sup_k N_k)^2} \; \right]\;   ,
$$
in agreement with Eq. (\ref{38}).
 
Another possibility could be to study the interplay between the correlation index for trapped 
atoms and the appearance of order parameters, coherence, compressibility, and other 
characteristics of Bose-condensed atoms. An interesting problem is the dependence of trapped 
atom properties on the trap-size variation \cite{Yukalov_2025}. Under this variation we mean 
a series of experiments with the traps of different sizes. For instance, turbulent motion in 
nonequilibrium Bose-Einstein condensates can display different scaling relations \cite{Moreno_57}
for different trap sizes.    
 
An important application could be the extension of the notion of correlation indices to the 
analysis of correlations in time series. The latter are ubiquitous in various branches of 
physics, physiology, economics, finance, etc. \cite{Small_58}. This could be possible 
resorting to the methods of space-state reconstruction from time series \cite{Casdagli_59} 
and reconstruction of complex networks induced by time series \cite{Zhang_60,Gao_62}. 
In the reconstructed state spaces or complex networks, it would be possible to define 
correlation functions connecting different points of state spaces or different network nodes,
after which the procedure, described in the paper, could be straightforwardly applicable.

\vskip 3mm

\section*{Funding}
 
No funding was received to assist with the preparation of this manuscript.

\section*{Financial interests}

The authors have no relevant financial or non-financial interests to disclose.

\section*{Competing Interests}

The authors have no competing interests to declare that are relevant to the content of
this article.

\section*{Authors' contribution}

All authors contributed to the study conception and design. Material preparation, data
collection and analysis were performed by V.I. Yukalov and E.P. Yukalova. Numerical 
analysis was accomplished by E.P. Yukalova. The first draft of the manuscript was 
written by V.I. Yukalov and all authors commented on previous versions of the manuscript. 
All authors read and approved the final manuscript.

\section*{Data availability} 

The authors declare that the data supporting the findings of this study are available within 
the paper, its supplementary information files, and the National Tibetan Plateau Data Center 
(https://doi.org/10.11888/Cryos.tpdc.272747).

\end{document}